\documentclass[amsmath, amsfonts, showpacs, twocolumn, prl]{revtex4}
\usepackage{graphicx}
\usepackage{epsfig}
\usepackage{bm}
\usepackage{euscript}
\usepackage{dcolumn}
\usepackage{color}
\usepackage{amsmath}
\usepackage{amssymb}

\begin{document}

\title{Coulomb drag at zero temperature}
\author{Alex Levchenko}
\author{Alex Kamenev}

\affiliation{Department of Physics, University of Minnesota,
Minneapolis, MN 55455, USA}

\begin{abstract}
We show that the Coulomb drag effect  exhibits saturation at small
temperatures, when calculated to the {\em third} order in the
interlayer interactions. The zero-temperature transresistance is of
the order  $h/(e^2 g^3)$, where $g$ is the dimensionless sheet
conductance. The effect is therefore the strongest in  low mobility
samples. This behavior should be contrasted with the conventional
(\textit{second} order) prediction that the transresistance scales
as a certain power of temperature and is (almost)
mobility-independent. The result demonstrates that the
zero-temperature drag  is {\em not} an unambiguous  signature of a
strongly-coupled state in double-layer systems.
\end{abstract}

\date{October 8, 2007}


 \maketitle


Coulomb drag effect has proven to be a sensitive probe of
electron-electron ({\em e-e}) interactions. The phenomenon is
usually observed
\cite{Solomon,Gramila,Sivan,Pillarisetty,Kellogg-1,Savchenko} in
double-layer systems, where electrons interact through the
long-ranged Coulomb forces. A current, passing  through one of the
layers (the active layer), induces  a voltage  across the second
(passive) layer. The ratio between the two, the so called drag {\em
transresistance} $\rho_{D}$, carries a valuable information about
the state of electrons in each of the layers, as well as the nature
of their mutual interactions.

The transresistance for  weakly interacting electrons was calculated
\cite{Laikhtman,Smith,MacDonald,Kamenev-Oreg,Flensberg} in the {\em
second order} in the screened interlayer interaction and found to be
given by
\begin{equation}\label{drag-known}
\rho_{D}(T)=0.12\,\frac{h}{e^{2}}
\left(\frac{T}{E_{F}}\right)^{2}\frac{1}{(\kappa d)^{2}(k_{F}d)^{2}}\, ,
\end{equation}
where $d$ is the separation between the layers, $E_{F}$ and $k_{F}$
are the Fermi energy and  momentum correspondingly and $\kappa$ is
the inverse Thomas-Fermi screening radius. This result is in a
reasonable agreement with a number of experiments
\cite{Solomon,Gramila,Sivan,Kellogg-1}. Its main feature is the {\em
quadratic} temperature dependence, which may be traced back to the
phase volume accessible  for the interlayer {\em e-e} scattering.
The second order  effect requires the electron-hole asymmetry, i.e.
the difference in velocity between electrons and holes on the
opposite sides of the Fermi surface. Such an asymmetry scales as
$E_F^{-1}$ for each of the two layers, giving rise to the factor
$E_F^{-2}$ in Eq.~(\ref{drag-known}). The latter serves as the
dimensional  scale, which normalizes the $T^2$ dependence.

On the other hand, the systems with strong interlayer correlations
are predicted to exhibit a {\em nonzero} drag transresistance
($\propto h/e^2$) even  at {\em zero temperature}. The examples
include 1D charge density waves at exact commensurability
\cite{Nazarov}, as well as Quantum Hall bilayer structures at the
total filling factor $\nu=1$ \cite{Stern}. In the latter system the
effect was likely observed experimentally in
Ref.~[\onlinecite{Kellogg-2}]. This raises a question if $\rho_D(0)$
may serve as an unambiguous indicator of a strongly-correlated state
in a system at hand. I.e. whether the drag transresistance undergoes
a quantum phase transition between the weakly-coupled state, where
it is strictly zero, and a strongly-coupled phase, where it is
finite.

In this Letter we give a strong argument against such a scenario. We
show that $\rho_D(0)\neq 0$ already in {\em weakly interacting}
bilayer systems. To this end we evaluate the transresistance in the
{\em third} order in the (screened) interlayer interactions and find
a constant temperature-independent contribution
\begin{equation}\label{drag-result}
\rho_{D}(T) = 0.27\, \frac{h}{e^{2}}\, \frac{1}{g^{3}}\,
\frac{1}{(\kappa d)^{2}}\, ; \quad \quad T<h/\tau\,,
\end{equation}
where $g=25.8k\Omega/\rho_{\square}$ is the dimensionless
conductance (here $\rho_{\square}$ is the resistance of the
single layer) and $\tau$ is the elastic scattering time.
Drag effect, saturating at small temperatures, is therefore
{\em not} an automatic indicator of a strongly correlated state.

There are general reasons to expect that the third order effect may
be qualitatively different from the second order one,
Eq.~(\ref{drag-known}). Indeed, the  third order transresistance
does {\em not} rely on the electron-hole asymmetry. This is because
the corresponding linear-response diagrams involve {\em four}-leg
vertices (see below) which do not vanish within linearized
dispersion relation approximation (i.e. in the electron-hole
symmetric case). Therefore, the result is expected to be independent
on the Fermi energy, $E_F$. Since we are interested in the lowest
temperatures, it is natural to focus on the diffusive  regime, where
$T\ll h/\tau$. In this regime there are no any other relevant
energy, which may provide a scale for a temperature dependence.
Hence, the temperature-independent result, Eq.~(\ref{drag-result}),
is not entirely unexpected. Moreover, the four-leg vertex, mentioned
above, is known to play a central role in the low-temperature
transport  of diffusive metals.  It is exactly this object that
gives rise to singular Altshuler-Aronov (AA) corrections to the {\em
intra}layer conductance \cite{AA}.

Coming from another perspective, it is certainly unusual to find a
temperature-independent result for the quantity which relies on the
{\em e-e} scattering rate. Indeed, the latter is proportional to the
available phase volume around the Fermi   surface, which  scales as
$T^2$. However, in addition to the occupation numbers the scattering
rate is proportional to a certain matrix element (the overlap
integral of six wave functions for the third order process,
considered here). In diffusive systems such matrix elements are
known to be singularly enhanced in the limit where all involved
states are close in energy \cite{Blanter-Mirlin}. It is exactly this
enhancement that leads to singular {\em e-e} interaction effects in
the low-temperature diffusive limit \cite{AA-review}. In the case of
the third order transconductance in 2D the smallness of the phase
volume is {\em exactly} compensated by the divergence of the
corresponding matrix elements. This yields  the temperature
independent transresistance, Eq.~(\ref{drag-result}). The diffusive
enhancement of the matrix elements is less pronounced in cleaner
systems.  Hence, in the clean limit, $g\to \infty$, the zero
temperature drag, Eq.~(\ref{drag-result}), disappears.

There are two limitations on the applicability of
Eq.~(\ref{drag-result}) at the very low temperatures. (i) Once the
temperature length $L_T=\sqrt{hD/T}$, where $D$ is the diffusion
constant, reaches the sample size $L$, the growth of the matrix
elements is saturated. As a result, $\rho_D\propto T^2$ at $T<E_{\rm
Th}$, here $E_{\rm Th}=hD/L^2$ is the Thouless energy. (ii) If the
sample size is very big, one may enter the regime of disorder and/or
interaction induced localization. The relevant temperature scale is
that where  AA correction $\sigma_{\square} =
e^2/h[g-\pi^{-1}\ln(h/T\tau)]$ \cite{AA} is significant, i.e. $T\sim
(h/\tau)e^{-\pi g}$. At smaller temperatures the diffusive
approximation breaks down and our result, Eq.~(\ref{drag-result}),
is not applicable.

A natural question  is why  in  experiments of e.g.
Refs.~[\onlinecite{Solomon,Gramila,Sivan}] the low-temperature
saturation of $\rho_{D}(T)$ was not observed. In order to answer,
one may estimate the saturation temperature $T^{*}$ by equating
Eqs.~\eqref{drag-known} and \eqref{drag-result}. This way, one finds
$T^{*}\approx E_{F} (k_F d)g^{-3/2}$. Employing the parameters of
e.g. Ref.~[\onlinecite{Gramila}]: $E_{F}\approx 60K$, $g\approx 100$
and $k_{F}d\approx 4$,  one finds $T^{*}\approx 0.25\mathrm{K}$ and
the residual resistance $\rho_{D}\approx0.4m\Omega$ as it follows
from the Eq.\eqref{drag-result}. At the same time the lowest
temperature reported in Ref.~[\onlinecite{Gramila}]
$T\approx0.5\mathrm{K}$ and the corresponding drag
$\rho_{D}\approx0.65m\Omega$ are just above the expected saturation.
The similar situation is true regarding most of the other reports of
the Coulomb drag Refs.~\cite{Solomon,Sivan,Kellogg-1,Savchenko}.

It is rather likely, though, that the saturation observed by Lilly
{\em et. al} Ref.~[\onlinecite{Lilly}] in  $\nu=1$ Quantum Hall
bilayer system in the composite fermion regime is a manifestation of
Eq.~(\ref{drag-result}). Indeed, it was shown \cite{Khveshchenko}
that the diffusive corrections in the composite fermion regime are
rather similar to those in zero magnetic field. Virtually the only
difference is a substantial downward renormalization of the
composite fermion conductance $g^{cf}$, as compared to the zero
field one, $g$. Estimating $g^{cf}\approx 10$ and
$E^{cf}_{F}\approx5\mathrm{K}$ for the samples of
Ref.~[\onlinecite{Lilly}], one finds $T^{*}\approx0.15\mathrm{K}$
and $\rho_D(0)\approx2\Omega$ in a good agreement with
Ref.~[\onlinecite{Lilly}]. To verify Eq.~(\ref{drag-result}) more
experiments in zero magnetic field with smaller $g$ or/and smaller
temperatures are needed.

\begin{figure}
\includegraphics[width=9cm]{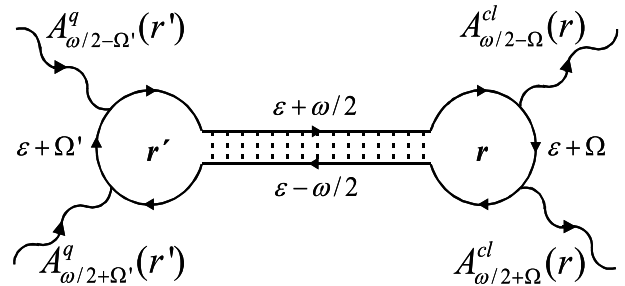}
\caption{The four-leg vertex, central to the third-order drag
effect, as well as to the intralayer  AA correction. External wavy
lines represent fluctuating vector potentials; the ladder is the
diffusion propagator $\mathcal{D}(r-r',\omega)$. \label{Fig-AA}}
\end{figure}

The four-leg vertex, which is a building block for  diagrams of the
{\em third} order drag transconductance is depicted in   Fig.~\ref{Fig-AA}.
It describes an induced  non-linear interaction of
electromagnetic fields through excitation of electron-hole pairs in
a given layer. The vertex is  non-local because of the diffusive propagation
of the electron-hole excitations within the layer.  The latter is encoded in the propagator
\begin{equation}\label{diffuson}
\mathcal{D}_\alpha(q,\omega)={1\over \nu_{\alpha}}\, \frac{1}{D_{\alpha}q^{2}-i\omega},
\end{equation}
where $\nu_{\alpha}$ is the density of states of the layer
$\alpha=1,2$  and $D_\alpha$ is its diffusion coefficient. Notice
that  the dimensionless  conductance is expressed as
$g_{\alpha}=\nu_{\alpha}D_{\alpha}$.

We  work with the Keldysh technique \cite{Keldysh,KA,Kamenev}. In
its framework the fluctuating electromagnetic potentials acquire  an
additional index: classical ($cl$) or quantum ($q$), which stay for
symmetric and antisymmetric combinations of the fields propagating
forward and backward in time, correspondingly. The proper indices
are indicated in Fig.~\ref{Fig-AA}. The fact that the four-leg
vertex of this very structure is {\em unique} in the leading order
in $1/g_\alpha$ may be rigorously proven within Keldysh non-linear
sigma-model \cite{KA}. In fact, it is exactly this vertex which
gives rise to the singular AA correction \cite{AA}. The latter is
obtained by pairing one classical and one quantum electromagnetic
potentials, while the two remaining ones represent an external
(classical) electric field along with induced current \cite{KA}.

It is convenient to work in a gauge, where the Coulomb interactions
are mediated by the longitudinal vector potentials, rather than the
scalar potentials. An advantage of using such a gauge is that both
internal and external potentials, as well as the current sources are
all expressed through the same type of field. This makes the
structure of the vertex, Fig.~\ref{Fig-AA}, particularly symmetric.
Moreover, the gauge may be chosen in a way that the propagator of
the longitudinal vector potentials
$\mathcal{V}_{\alpha\beta}=2i\langle A^{cl}_\alpha A^{q}_\beta
\rangle$ automatically includes the vertex renormalization  by the
disorder  \cite{KA}
\begin{equation}
\mathcal{V}_{\alpha\beta}(q,\omega)=
\frac{q^{2}V^{R}_{\alpha\beta}(q,\omega)}
{(D_{\alpha}q^{2}-i\omega)(D_{\beta}q^{2}-i\omega)}
\, ,
                                                              \label{Contractions}
\end{equation}
where $V^R_{\alpha\beta}(q,\omega)$ is the $2\times2$ matrix of
retarded intra and interlayer interactions calculated within random
phase approximation (RPA). The latter is the solution of the
standard Dyson equation \cite{Kamenev-Oreg,Flensberg}
$\hat{V}^R=\hat{V}_{0}+\hat{V}_{0}\hat{\Pi}\hat{V}^R$, where
\begin{equation}
                               \label{RPA}
\hat{V}_{0}\!=\!\frac{2\pi e^{2}}{q}\!\left(\begin{array}{cc} 1 & e^{-qd}
\\ e^{-qd} & 1
\end{array}\right),\,\,\,\,
\hat{\Pi}\!=\!\left(\begin{array}{cc}
\frac{\nu_{1}D_{1}q^{2}}
{D_{1}q^{2}-i\omega} & 0 \\ 0 &
\frac{\nu_{2}D_{2}q^{2}}
{D_{2}q^{2}-i\omega}\end{array}\right).
\end{equation}
Note that the polarization operator $\hat{\Pi}(q,\omega)$ has
no off-diagonal elements, reflecting the
absence of tunneling between the layers.

\begin{figure}
\includegraphics[width=9cm]{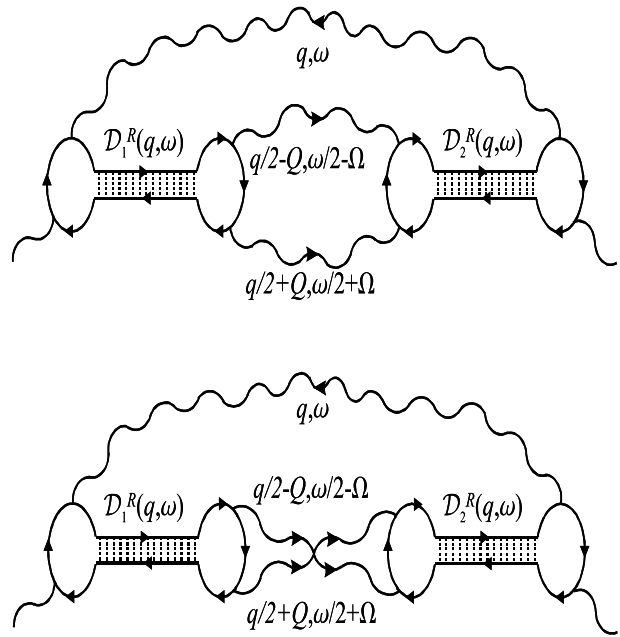}
\caption{Two diagrams for the drag transconductance $\sigma_{D}$ in
the third order in the interlayer interactions,
$\mathcal{V}_{12}(q,\omega)$, denoted by  wavy lines. The intralayer
diffusion propagators $\mathcal{D}_{\alpha}(q,\omega)$,
Eq.~(\ref{diffuson}), are denoted by ladders.\label{Fig-Drag}}
\end{figure}

We are now on the position to evaluate the third order drag
transconductance. The corresponding diagrams are constructed from
the two vertices of Fig.~\ref{Fig-AA}: one for each of the layers,
Fig.~(\ref{Fig-Drag}). Remarkably, there are only two ways to
connect them, using the propagators (\ref{Contractions}) (recall
that $\langle A^{q}_\alpha A^{q}_\beta \rangle=0$, \cite{Kamenev}).
The analytic expression for the sum of the two diagrams of
Fig.~\ref{Fig-Drag}  is given by
\begin{eqnarray}\label{sigma-general}
&& \!\!\!\!\!\! \sigma_{D}\!=\!32e^{2}Tg^{2}_{1}g^{2}_{2}\int\limits^{\infty}_{0}
\frac{\mathrm{d}\omega\mathrm{d}\Omega}{4\pi^{2}}\,
\mathcal{F}_{1}\mathcal{F}_{2}\!
\sum_{q,Q}\mathrm{Im}\Big[\mathcal{D}_{1}(q,\omega)\mathcal{D}_{2}(q,\omega)
\nonumber \\
&& \!\!\!\!\!\! \mathcal{V}_{12}(q,\omega)
\mathcal{V}_{12}\left({q\over 2}-Q,{\omega\over 2}-\Omega\right)
\mathcal{V}_{12}\left({q\over 2}+Q,{\omega\over 2}+\Omega\right)\Big].
\end{eqnarray}
The two functions $\mathcal{F}_{1}(\omega,\Omega)$ and
$\mathcal{F}_{2}(\omega,\Omega)$ originate from the integration over
the fast electronic energy $\varepsilon$, Fig.~\ref{Fig-AA}, in the
active and passive layers correspondingly. In the dc limit they are
given by
\begin{subequations}\label{Spectral-F}
\begin{equation}
\mathcal{F}_{1}(\omega,\Omega)=T\frac{\partial}{\partial\Omega}
\left[\mathcal{B}(\Omega+\omega/2)-\mathcal{B}(\Omega-\omega/2)\right],
\end{equation}
\begin{equation}
\mathcal{F}_{2}(\omega,\Omega)=2-\mathcal{B}(\Omega+\omega/2)
-\mathcal{B}(\Omega-\omega/2)+\mathcal{B}(\omega),
\end{equation}
\begin{equation}
\mathcal{B}(\omega)=\frac{\omega}{T}\coth\left(\frac{\omega}{2T}\right).
\end{equation}
\end{subequations}

To make the farther calculations more compact, we restrict ourselves
to the identical layers. We shall first consider the experimentally
most relevant case of the long-ranged coupling, where $\kappa d\gg
1$. Here $\kappa=2\pi e^2\nu$ is the Thomas-Fermi inverse screening
radius. In this limit the effective interlayer interaction
potential, Eqs.~(\ref{Contractions}), (\ref{RPA}), acquires a simple
form
\begin{equation}
\mathcal{V}_{12}(q,\omega)=\frac{1}{g}\, \frac{1}{\kappa d
Dq^{2}-2i\omega}\,.
                           \label{V12}
\end{equation}
Next, we substitute Eqs.~(\ref{diffuson}), (\ref{Spectral-F}) and
(\ref{V12}) into Eq.~(\ref{sigma-general}) and perform the energy
and momentum integrations. The inspection of the integrals shows
that both energies $\omega$ and $\Omega$ are of the order of the
temperature $\omega\sim\Omega\sim T$ (in compliance with the phase
volume considerations) \cite{foot1}. On the other hand, the
characteristic value   of the transferred momenta is $q\sim Q\sim
\sqrt{T/(D\kappa d)}\ll \sqrt{T/D}$, cf. Eq.~(\ref{V12}). Therefore
one may disregard $Dq^2$ as compared to $i\omega$ in the expressions
for ${\mathcal D}_{\alpha}(q,\omega)$,  Eq.~(\ref{diffuson}),
approximating the product ${\mathcal D}_1{\mathcal D}_2$ in
Eq.~(\ref{sigma-general}) by $-\omega^{-2}$. This factor represents
the diffusive enhancement of the matrix elements, mentioned in the
introduction. Such spatial scales separation implies that the
four-leg vertices, Fig.~\ref{Fig-AA}, are effectively spatially
local, while the three interlayer interaction lines are long-ranged.

Rescaling  energies by $T$ and momenta by $\sqrt{T/(D\kappa d)}$,
one may reduce the expression (\ref{sigma-general}) for the
transconductance to $\sigma_D=(e^2/h)\,g^{-1}(\kappa d)^{-2}\times $
[{\rm dimensionless integral}]. The latter integral does not contain
any parameters and is free from divergences in all directions. It is
thus simply a number that may be evaluated numerically
\cite{Coefficient}. In the limit $\sigma_D\ll (e^2/h)\, g_\alpha$
the transresistance is related to $\sigma_D$ by $\rho_D=
\sigma_Dh^2/(e^4 g_1g_2)$, resulting  finally in
Eq.~(\ref{drag-result}).

To emphasize the fact that the scale separation, discussed above, is
{\em not} crucial for having the low temperature saturation, we
briefly consider the case of the short-ranged interlayer
interactions, $V^R_{12}(q,\omega)=V_0$. The latter may be relevant,
if interactions are screened by e.g. metallic  back gate. One
employs then Eqs.~(\ref{diffuson}), (\ref{Contractions}) and
rescales  the energies by the temperature, while the momenta by
$\sqrt{T/D}$. This way the transconductance,
Eq.~(\ref{sigma-general}), once again reduces to the dimensionless
and parameter free integral. The latter is convergent in all
directions and may be readily evaluated, resulting in
 \begin{equation}\label{short-ranged}
    \rho_D=0.01\ {h\over e^2}\, {1\over g^3}\, (\nu V_0)^3\, .
 \end{equation}
Notice, that the effect is expected to have the negative sign for
the short-ranged {\em attractive} interactions. This observation may
have relevance for oppositely doped double layer structures.

The low temperature saturation of the Coulomb drag was discussed
previously in Refs.~[\onlinecite{Gornyi}] and [\onlinecite{Zhou}].
Both of them considered essentially  different and somewhat more
exotic mechanisms. The zero temperature saturation suggested in
Ref.~[\onlinecite{Gornyi}] relies on the assumption that  the
electrons in both layers are scattered by exactly the same disorder
potential. Ref.~[\onlinecite{Zhou}] focuses on the strongly coupled
regime, where the pairing order parameter is suppressed by disorder.

To conclude, we studied the Coulomb drag phenomenon in weakly
interacting bilayer systems. We found that effect saturates at small
temperatures, when calculated to the third order in the interlayer
interactions. The saturation of drag relies on the presence of
disorder and scales inversely with mobility. It does not require,
though, any correlations of the disorder potential in the two
layers. The effect was possibly observed in
Ref.~[\onlinecite{Lilly}], although more experiments in lower
mobility samples and zero magnetic field are highly desirable.

We are grateful  to D.~Bagrets, L.~Glazman, I.~Gornyi, F.~von Oppen,
A.~Savchenko, B.~Shklovskii, A.~Stern for stimulating discussions.
This work was supported by NSF Grant No. DMR 0405212. A.K. is also
supported by the A.~P.~Sloan foundation.



\begin{thebibliography}{99}
\bibitem{foot1} This behavior should be contrasted with  AA effect
\cite{AA,AA-review}, where the frequency integral comes from the
range $T\lesssim\omega\lesssim h/\tau$.
\bibitem{Coefficient} The evaluation of the integrals over momenta is
substantially simplified by the local nature of the vertex (in the limit
$\kappa d\gg 1$). Transforming to the real space representation,
the 4-fold momentum integral in Eq.~(\ref{sigma-general}) reduces to
$$\int^{\infty}_{0}\mathrm{d}r\
r\ \mathrm{Im}[K_{0}(r\mu)K_{0}(r\mu_{+})K_{0}(r\mu_{-})]=M(x,y),$$
where $K_0(r\mu)$ is the modified Bessel function, which is the 2D
Fourier transform of the interaction potential (\ref{V12}). Here
$\mu=\sqrt{-\mathrm{i}x}$ and $\mu_{\pm}=\sqrt{-\mathrm{i}(\pm
y+x/2)}$, the radius is normalized by $\sqrt{D\kappa d/(2T)}$ and
$x=\omega/T$; $y=\Omega/T$. Finally the number of interest is given
by
$$-2\pi^{-3}\iint\limits^{\quad\infty}_{0}\mathrm{d}x\mathrm{d}y
\mathcal{F}_{1}(x,y)\mathcal{F}_{2}(x,y)x^{-2}M(x,y)\approx 0.27$$
\end{thebibliography}
\end{document}